\documentclass[fleqn,usenatbib]{mnras}

\usepackage{newtxtext,newtxmath}

\usepackage[T1]{fontenc}
\usepackage{ae,aecompl}

\usepackage{graphicx}	
\usepackage{amsmath}	
\usepackage{amssymb}	

\newcommand{\kmps}{\ensuremath{\mathrm{km~s^{-1}}}}
\newcommand{\Msun}{\ensuremath{\mathrm{M_\odot}}}

\newcommand{\Msunpyr}{\ensuremath{\mathrm{M_\odot~yr^{-1}}}}
\newcommand{\Mej}{\ensuremath{M_\mathrm{\mathrm{ej}}}}
\newcommand{\Eej}{\ensuremath{E_\mathrm{\mathrm{ej}}}}
\newcommand{\Ni}{\ensuremath{\mathrm{^{56}Ni}}}

\title[AIC in M81?]{
VTC~J095517.5+690813: A radio transient from an accretion-induced collapse of a white dwarf?
}

\author[T. J. Moriya]{
Takashi J. Moriya\thanks{E-mail: takashi.moriya@nao.ac.jp}
\\
National Astronomical Observatory of Japan, National Institutes of Natural Sciences, 2-21-1 Osawa, Mitaka, Tokyo 181-8588, Japan\\
School of Physics and Astronomy, Faculty of Science, Monash University, Clayton, VIC 3800, Australia
}

\date{Accepted 2019 September 14. Received 2019 September 13; in original form 2019 August 15}

\pubyear{2019}

\begin{document}
\label{firstpage}
\pagerange{\pageref{firstpage}--\pageref{lastpage}}
\maketitle

\begin{abstract}
We investigate a possibility that a recently reported radio transient in M81, VTC~J095517.5+690813, is caused by an accretion-induced collapse of a white dwarf. It became bright in radio but no associated optical transient was discovered. The accretion-induced collapse is predicted to be radio bright but optically faint, satisfying the observed property. We compare predicted radio emission from the accretion-induced collapse with that of VTC~J095517.5+690813 and show that it can be reasonably explained by the accretion-induced collapse. Although it is difficult to firmly conclude that VTC~J095517.5+690813 is an accretion-induced collapse, our study shows that radio-bright transients without an optical counterpart could still be related to stellar deaths.
\end{abstract}




\section{Introduction}
Time domain astronomy is a current frontier of astronomy. The variable sky has long been investigated in optical but transient surveys in non-optical wavelengths are growing recently \citep[e.g.,][]{mooley2016radiosurvey,kasliwal2017spiritz}. For example, fast radio bursts (FRBs) that show luminosity variations in radio in the timescale of milliseconds clearly shows that the sky is very dynamic in time in the non-optical wavelengths \citep{lorimer2007frb,thornton2013frb}.

\citet{anderson2019m81rad} recently reported a radio transient discovered by their radio transient survey with Karl G. Jansky Very Large Array. The transient, named VTC~J095517.5+690813, was discovered in 6~GHz in M81 (3.6~Mpc). No optical transients in M81 were reported so far at the location \citep{anderson2019m81rad}. If bright optical transients were accompanied, the modern transient surveys such as ASAS-SN\footnote{\url{http://www.astronomy.ohio-state.edu/~assassin/}} would have been discovered them because of its proximity. Therefore, VTC~J095517.5+690813 was likely not associated with a bright optical transient. X-ray observations at the transient location were conducted before and after the discovery of VTC~J095517.5+690813 but no X-ray source was found at the transient location, either. The pre-transient images provided by the \textit{Hubble Space Telescope} (\textit{HST}) found two red giants at the location of VTC~J095517.5+690813, which could be a chance coincidence, and no luminous stars were present at the location. Therefore, the origin of VTC~J095517.5+690813 is not likely related to deaths of massive stars during which optically-faint radio-bright transients may be generated \citep[e.g.,][]{kashiyama2018bhradio}. The exact nature of VTC~J095517.5+690813 has not been clarified yet.

Accretion-induced collapse (AIC) is a theoretically-predicted collapse of a white dwarf (WD) forming a neutron star (NS). If an O+Ne+Mg WD grows its mass close to the Chandrasekhar mass limit, electron-capture reactions can be initiated at the center \citep[e.g.,][]{nomoto1991aic}. The electron-degeneracy pressure supporting the WD is suddenly lost due to the electron-capture reactions and the WD collapses until it becomes a NS \citep[e.g.,][]{woosley1992aic}. Because AIC results in a formation of a NS from a WD, it has been suggested that young NSs in globular clusters and some millisecond pulsars are formed through AIC \citep[e.g.,][]{bailyn1990aic,bhattacharya1991aic,tauris2013aic}.

Identifying AIC is challenging. Numerical simulations of AIC predicts little ($\sim 0.0001~\Msun$, \citealt{dessart2006aic,dessart2007aic}) or small ($\sim 0.01~\Msun$, \citealt{metzger2009aic,darbha2010aic}) amount of \Ni\ production and AIC is not likely accompanied by bright optical transients (but see also \citealt{piro2014rloaicopt}). However, it has been suggested that AIC can result in bright radio transients \citep[e.g.,][]{piro2013aicradio,moriya2016aic}. Because the lack of the optical counterpart as well as the faintness of the progenitor found in VTC~J095517.5+690813 are consistent with the expected properties of AIC, we investigate the possibility that VTC~J095517.5+690813 results from AIC in this paper.

\section{Method}
We consider the radio emission from AIC resulting from the interaction between the forward shock and circumstellar matter (CSM) as in \citet{moriya2016aic}. Relativistic electrons accelerated at the forward shock emits synchrotron radiation that can be observed in radio. The spectral index of VTC~J095517.5+690813 is not well constrained but it is consistent with that expected from synchrotron emission \citep{anderson2019m81rad}. The ejecta properties from AIC and the CSM properties at AIC determine the radio properties of AIC. In this section, we first summarize the predicted ejecta and CSM properties of AIC. Then, we present our radio emission model.

\subsection{AIC properties}
\subsubsection{Ejecta properties}
Ejecta from AIC are theoretically suggested to have several different properties. We take three possible combinations of ejecta mass (\Mej) and ejecta kinetic energy (\Eej) in this work as summarized in Table~\ref{tab:aicprop}.

The multi-dimensional explosion simulations of AIC conducted by \citet{dessart2006aic} found that the neutrino-driven explosions of AIC lead to $\Mej\sim 0.001~\Msun$ and $\Eej\sim 10^{49}~\mathrm{erg}$ (Model~A). A later work by the same group \citep{dessart2007aic} which added the magnetohydrodynamic (MHD) effect in the explosion simulations found that the magnetic stress can significantly change the AIC ejecta properties, i.e., $\Mej\sim 0.1~\Msun$ and $\Eej\sim 10^{51}~\mathrm{erg}$ (Model~B). In both cases, the amount of \Ni\ synthesized during the AIC explosions ($M_\mathrm{^{56}Ni}$) is only $\sim 0.0001~\Msun$ and, therefore, no optical transients powered by the radioactive decay are likely accompanied by AIC.

On the other hand, \citet{metzger2009aic} and \citet{darbha2010aic} suggested that a \Ni-rich outflow with $\Mej\sim M_\mathrm{^{56}Ni} \sim 0.01~\Msun$ may be ejected if a rapidly rotation WD causes AIC. An neutron-rich accretion disk can be formed when the WD collapses. Although the disk is initially neutron-rich, the neutrino emission from the collapsing WD would make the proton-to-neutron ratio in the disk to be about 1. Thus, the composition of the hot accretion disk becomes dominated by \Ni. The disk can be ejected thanks to the viscous stress and the nuclear fusion energy. They extimate the typical ejecta velocity of $\sim 0.1c$, where $c$ is the speed of light, and therefore $\Eej\sim 10^{50}~\mathrm{erg}$ (Model~C).

\begin{table}
	\centering
	\caption{Predicted AIC ejecta properties.}
	\label{tab:aicprop}
	\begin{tabular}{lcccl} 
		\hline
		Model & $M_\mathrm{ej}$ & $E_\mathrm{ej}$ & $M_\mathrm{^{56}Ni}$ & Reference\\
	 & $M_\odot$ & $10^{51}~\mathrm{erg}$ & $M_\odot$ & \\
		\hline
		A & $\sim 0.001$ & $\sim 0.01$ & $\sim 0.0001$ & \citet{dessart2006aic} \\
		B & $\sim 0.1$ & $\sim 1$ & $\sim 0.0001$ & \citet{dessart2007aic} \\
		C & $\sim 0.01$ & $\sim 0.1$ & $\sim 0.01$ & \citet{metzger2009aic} \\
		\hline
	\end{tabular}
\end{table}

\subsubsection{CSM properties}\label{sec:CSMporp}
There are two major channels to cause AIC. One is through the accretion onto an O+Ne+Mg WD by a non-degenerate companion star (single-degenerate [SD] channel) and the other is by a merger of two C+O WDs (double-degenerate [DD] channel). The two channels predict different CSM properties and we summarize them in this section.

When the mass of the O+Ne+Mg WD in the SD channel grows to the Chandrasekhar mass limit, the WD causes AIC. The accretion rate to grow the WD mass is estimated to be $\sim 10^{-7}-10^{-5}~\Msunpyr$ \citep[e.g.,][]{nomoto1991aic,nomoto2007aic,shen2007wdacc,wang2017heaic}. The accretion is provided by the non-degenerate companion star by binary mass transfer process. Some material is ejected from the binary system during the mass transfer process that can form a CSM. If the mass-loss rate from the system is $\dot{M}_\mathrm{loss}$ and the CSM velocity is $v_\mathrm{CSM}$, the CSM density $\rho_\mathrm{CSM}$ is expressed as
\begin{equation}
    \rho_\mathrm{CSM}(r) = \frac{\dot{M}_\mathrm{loss}}{4\pi v_\mathrm{CSM}}r^{-2},
\end{equation}
from the mass conservation. Following the convention, we define $4\pi A_\star = (\dot{M}_\mathrm{loss}/v_\mathrm{CSM})/(10^{-5}~\Msunpyr/1000~\kmps)$ so that the CSM density profile can be expressed as $\rho_\mathrm{CSM}(r)=5\times 10^{11} A_\star r^{-2}$ (cgs unit). The CSM density is solely determined by $A_\star$ and we change $A_\star$ in our radio light-curve (LC) modeling. As summarized in \citet{moriya2016aic}, $A_\star\sim 0.001-10$ is predicted from the SD channel of AIC.

The other path to AIC is through a merger of two C+O WDs (the DD chanel). During the merger of the WDs, one WD can be tidally disrupted and the material from the disrupted WD can accrete onto the other WD \citep[e.g.,][]{dan2014wdmerger,shen2015wdmerger,sato2016wdmerger}. As a result, the off-center C burning can be ignited at the surface of the survived WD and the C+O WD can be transformed to an O+Ne+Mg WD as the burning front proceeds to the center. If the newly-formed O+Ne+Mg WD is more massive than the Chandrasekhar mass limit, the WD can collapse and cause AIC when its central density becomes high enough after it cools down \citep[e.g.,][]{yoon2005massivewd}. Because little mass loss is expected from the WD binary and there is a time delay from the merger to AIC, the CSM density around AIC from the DD channel is not likely to be much different from the interstellar matter (ISM) density. Thus, we assume that the CSM density around AIC from the DD channel is constant with the canonical ISM density of around $1~\mathrm{cm^{-3}}$ \citep{ferriere2001ism}.
We note, however, \citet{schwab2016supc} recently showed that the WD-WD merger product may inflate to form a giant star having a large mass-loss rate. Although the large mass-loss rate does not continue until the time of AIC, the CSM density may not be as low as that of ISM in this case. A shell-like CSM may also exist in such a case.

\subsection{Radio emission model}
The synchrotron luminosity at the frequency $\nu$ ($L_\nu$) from the forward shock can be formulated as \citep{fransson1998sync,bjornsson2004sync}
\begin{equation}
\nu L_\nu \simeq \pi r_\mathrm{sh}^2 v_\mathrm{sh} n_\mathrm{rel}
\left(\frac{\gamma_\nu}{\gamma_\mathrm{min}}\right)^{1-p}
\gamma_\nu m_e c^2
\left[
1+\frac{t_{\mathrm{sync},\nu}}{t}
\right]^{-1},
\label{eq:radiolum}
\end{equation}
where $r_\mathrm{sh}$ is the forward shock radius, $v_\mathrm{sh}$ is the forward shock velocity, $n_\mathrm{rel}$ is the number density of the accelerated electrons, $\gamma_\nu=(2\pi m_ec\nu/eB)^{0.5}$ is the Lorentz factor of the accelerated electrons with the characteristic frequency $\nu$, $\gamma_\mathrm{min}$ is the minimum Lorentz factor of the accelerated electrons, $m_e$ is the electron mass, $t_\mathrm{sync,\nu}$ is the synchrotron cooling timescale, $e$ is the electron charge, and $B$ is the magnetic field strength. We assume $\gamma_\mathrm{min}\sim 1$ because the typical shock velocity is of the order of $0.1c$. We also assume that the number density of electrons accelerated to the Lorentz factor of $\gamma$ is proportional to $\gamma^{-p}$ with $p=3$. The synchrotron cooling timescale
$t_\mathrm{sync,\nu}$ is $6\pi m_e c/\sigma_T\gamma_\nu B^2$, where $\sigma_T$ is the Thomson scattering cross section.

A fraction $\varepsilon_B$ of the shock kinetic energy is assumed to be converted to the magnetic field energy and a fraction $\varepsilon_e$ is assumed to be used for the electron acceleration at the shock in estimating the synchrotron luminosity. These fractions are quite uncertain. Detailed analysis of radio and X-ray emission from SN shocks show that $\varepsilon_B\sim 0.01-0.001$ and $\varepsilon_e\sim 0.1$ \citep[e.g.,][]{bjornsson2004sync,maeda2012sn2011dhrx,kamble2016sn2013dfrx}. We here assume $\varepsilon_B = 0.01$ and $\varepsilon_e = 0.1$.

The synchrotron emission formulated above is altered by absorption processes. The major absorption process is the synchrotron self-absorption (SSA) at the shock \citep{chevalier1998ssa}, because the CSM density of our interest is low enough to neglect the free-free absorption by the unshocked CSM. The SSA optical depth of $\tau_\mathrm{SSA}=(\nu/\nu_\mathrm{SSA})^{-(p+4)/2}$ is adopted, where $\nu_\mathrm{SSA}\simeq 3\times 10^{5}(r_\mathrm{sh}\epsilon_e/\epsilon_B)^{2/7}B^{9/7}$~Hz in the cgs unit for $p=3$.

Finally, the forward shock properties ($r_\mathrm{sh}$ and $v_\mathrm{sh}$) are derived by using the self-similar solution of \citet{chevalier1982selfsimilar}. For this purpose, the AIC ejecta are assumed to have the double power-law density profile ($\propto r^{-n}$ outside and $\propto r^{-\delta}$ inside) with $n=10$ and $\delta =1$ as in supernova ejecta from compact progenitors \citep{matzner1999mckee}.

\begin{figure}
\includegraphics[width=0.95\columnwidth]{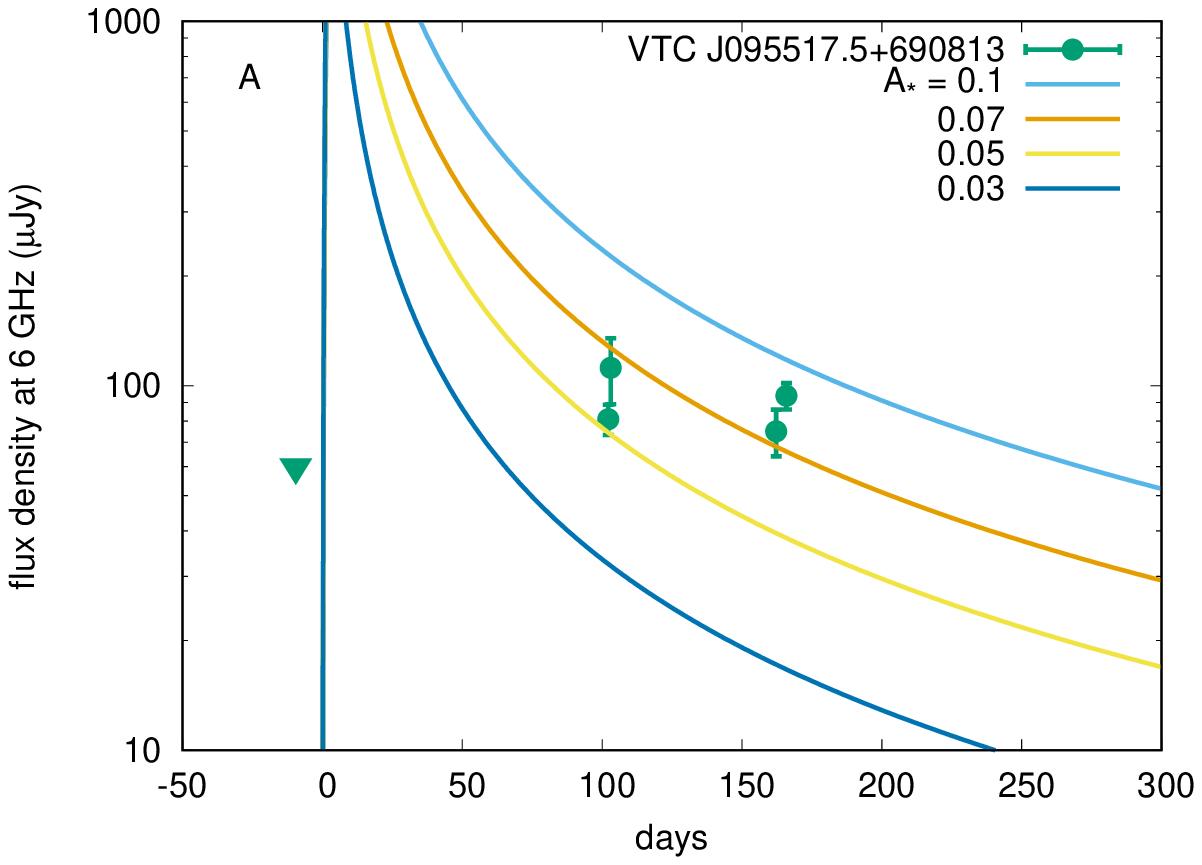}
\includegraphics[width=0.95\columnwidth]{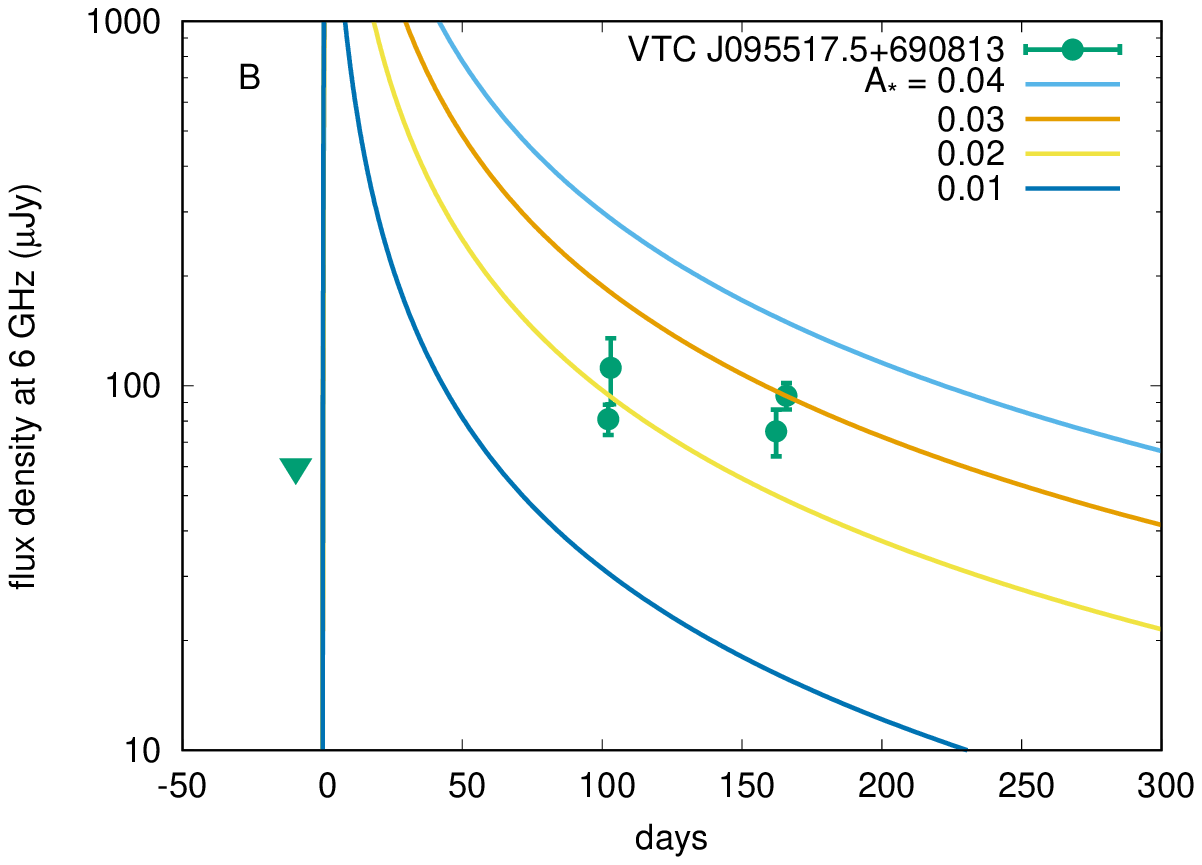}
\includegraphics[width=0.95\columnwidth]{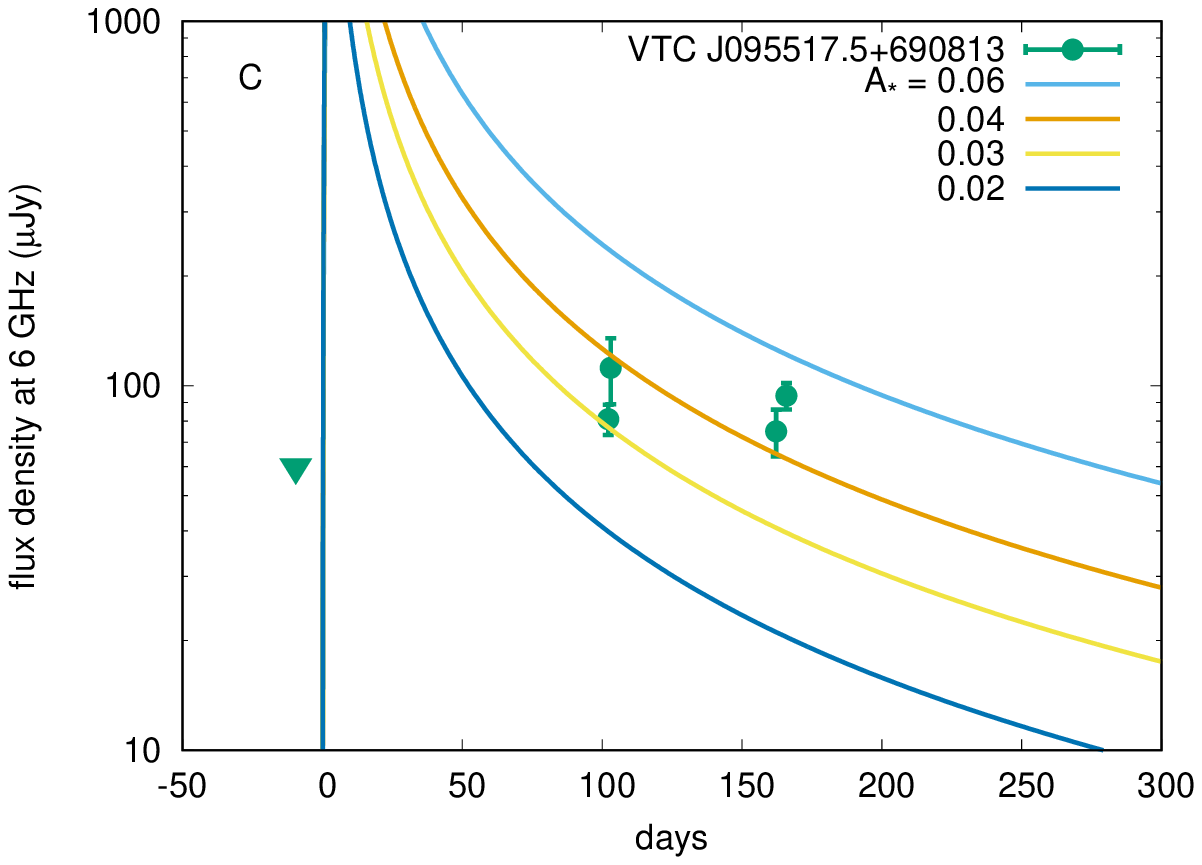}
\caption{
Radio (6~GHz) LC of AIC from the SD channel having the CSM density profile of $\rho_\mathrm{CSM}(r)=5\times 10^{11}A_\star r^{-2}$ (cgs unit) at the distance of VTC~J095517.5+690813 (3.6~Mpc). The different panels have the different ejecta properties (Table~\ref{tab:aicprop}). The radio LC of VTC~J095517.5+690813 at 6~GHz \citep{anderson2019m81rad} is presented for comparison. The triangle shows the upper flux limit. The explosion date of the synthetic LCs is 0~days and the observed LC is shifted to compare with the synthetic LCs.
}
\label{fig:lightcurve_wind}
\end{figure}

\begin{figure}
\includegraphics[width=0.95\columnwidth]{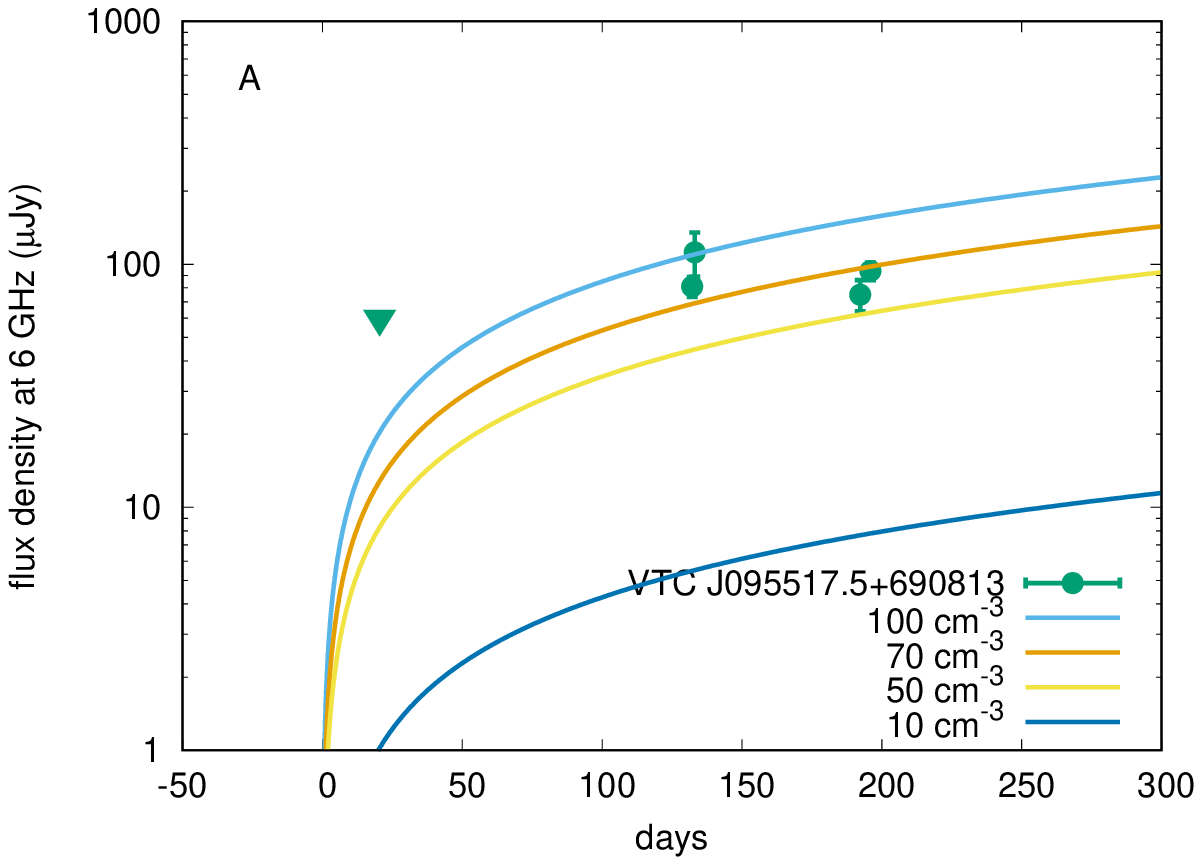}
\includegraphics[width=0.95\columnwidth]{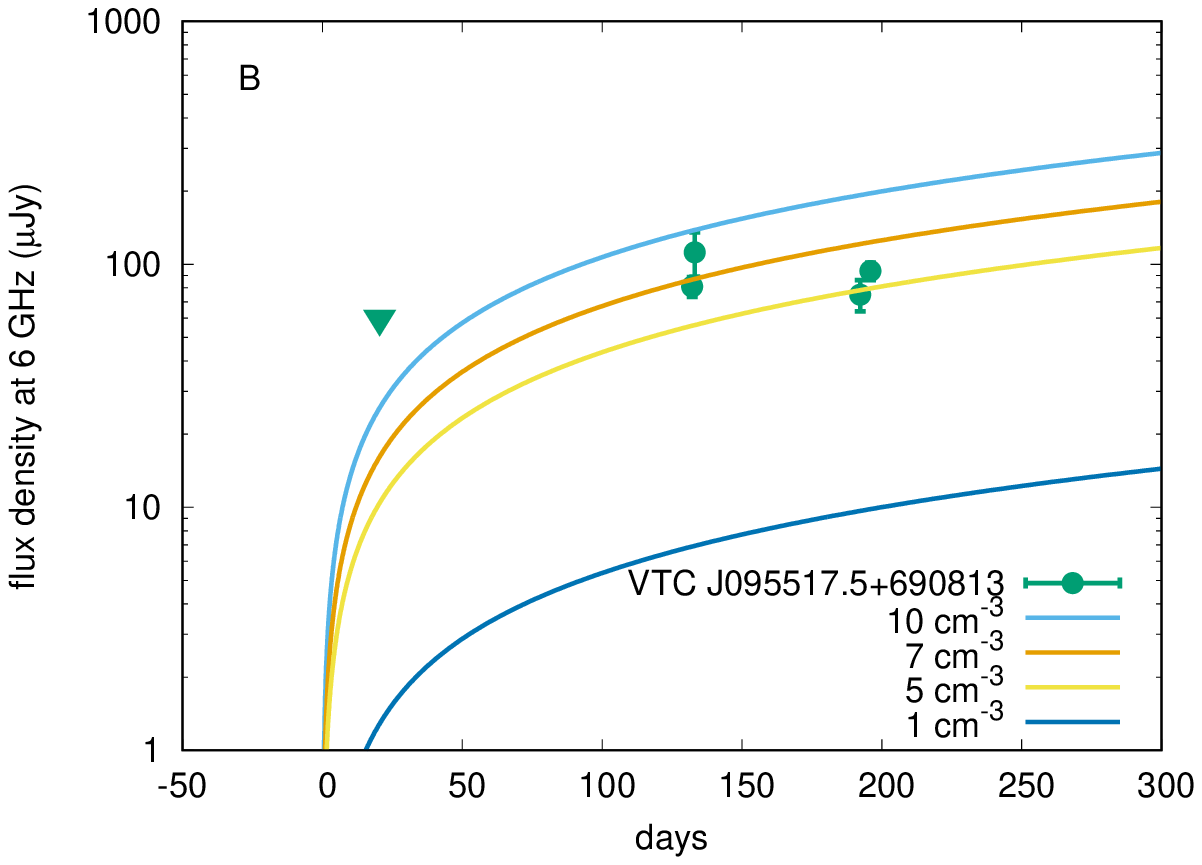}
\includegraphics[width=0.95\columnwidth]{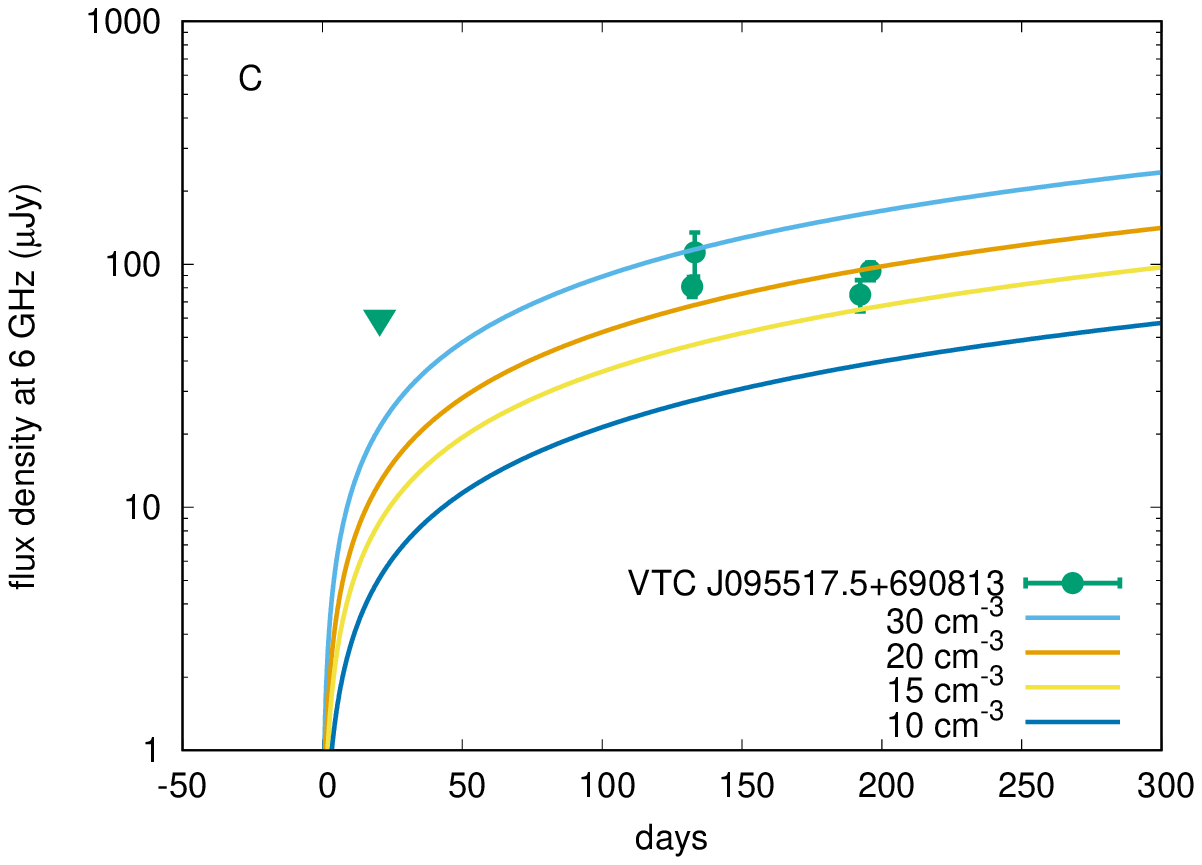}
\caption{
Radio (6~GHz) LC of AIC from the DD channel having the constant CSM density at the distance of VTC~J095517.5+690813 (3.6~Mpc). The different panels have the different ejecta properties (Table~\ref{tab:aicprop}). The radio LC of VTC~J095517.5+690813 at 6~GHz \citep{anderson2019m81rad} is presented for comparison. The triangle shows the upper flux limit. The explosion date of the synthetic LCs is 0~days and the observed LC is shifted to compare with the synthetic LCs.
}
\label{fig:lightcurve_const}
\end{figure}

\section{AIC model for VTC~J095517.5+690813}
We analyse VTC~J095517.5+690813 in the context of AIC in this section. The SD and DD channels have different CSM density structure and we treat them separately.

\subsection{SD channel}
The CSM around AIC from the SD channel has the density structure of $\rho_\mathrm{CSM} (r)=5\times 10^{11}A_\star r^{-2}$ (cgs unit). Fig.~\ref{fig:lightcurve_wind} shows the synthetic radio (6~GHz) LCs of AIC from the SD channel with different $A_\star$ and compares them with the radio LC of VTC~J095517.5+690813 observed at 6~GHz. The synthetic radio LCs first show a rapid rise time of less than 10~days, which corresponds to the time in which the SSA optical depth becomes unity at 6~GHz. Then the synthetic radio LCs decline following the density decline in CSM. The radio LC of VTC~J095517.5+690813 is consistent with the radio luminosity evolution of AIC at the declining phase with $A_\star\simeq 0.01-0.1$.

The required CSM density $A_\star\simeq 0.01-0.1$ is relatively low. The low CSM density of $A_\star\lesssim 0.1$ is expected in the SD system with a stable nuclear burning phase with a main-sequence or He star companion. If red giants are companion, the slow expected wind velocity ($\sim 10~\kmps$) makes the CSM density large ($A_\star \simeq 0.1-10$), despite of their small mass-loss rates ($\sim 10^{-8}-10^{-6}~\Msunpyr$). Thus, the red giants discovered at the location of VTC~J095517.5+690813 in \textit{HST} pre-transient images are not likely a donor star leading a WD to AIC. This is also consistent with the lack of optical transients, because it is suggested that AIC led by extended donors can become bright in optical even if little \Ni\ is produced \citep{piro2014rloaicopt}. Several studies suggest that the He star donor channel could be a major path leading to AIC \citep[e.g.,][]{wang2017heaic,wang2018aic,brooks2017aiche,ruiter2019aic}. The He star donor channel for AIC experience the optically-thick wind phase in which $A_\star \gtrsim 0.1$ is expected \citep{wang2017heaic} but some systems can come down to the stable burning phase before AIC \citep{wang2018aic}.

\subsection{DD channel}
The CSM around AIC from the DD channel is expected to have a constant density similar to the ISM density (but see also \citealt{schwab2016supc}). Fig.~\ref{fig:lightcurve_const} shows our synthetic radio LC models in the case of the constant CSM density and compares them with that of VTC~J095517.5+690813 in 6~GHz. Contrary to the SD channel case, the radio LCs keep increasing. This is because the CSM density is constant and also the SSA is not effective here because of the low CSM density. The radio luminosity keeps increasing as the shock propagates outward during the epochs shown in Fig.~\ref{fig:lightcurve_const}. 

As presented in Fig.~\ref{fig:lightcurve_const}, the DD channel models can also provide a good match to VTC~J095517.5+690813. The CSM density to account for VTC~J095517.5+690813 varies significantly depending on the assumed ejecta properties. For Model~A, a high ISM density of around $70~\mathrm{cm^{-3}}$ is required to account for the luminosity of VTC~J095517.5+690813. In Model~B, about $7~\mathrm{cm^{-3}}$ is required to explain VTC~J095517.5+690813 and the estimated density is consistent with that of ISM. Model~C requires a rather large CSM density of around $20~\mathrm{cm^{-3}}$. Overall, the radio LC of VTC~J095517.5+690813 can be explained by AIC from the DD channel, but the MHD-driven explosion ejecta model (Model~B) is preferred because of its low CSM density.

\section{Discussion}
\subsection{AIC or not?}
We have shown that the radio LC of VTC~J095517.5+690813 is consistent with the synthetic radio LCs of AIC. However, it is difficult to firmly conclude that VTC~J095517.5+690813 is indeed an AIC. This is partly because of the limited LC information available for VTC~J095517.5+690813. A longer monitoring of the radio LC evolution is required to identify radio transients from AIC conclusively.

Despite of the expected optical faintness, the multi-wavelength follow-up of radio transients would be important to identify AIC. The forward shock as well as the possible pulsar wind nebula formation can make AIC not only bright in radio but also in X-ray \citep{yu2019aicxray}. VTC~J095517.5+690813 was observed in X-ray but no X-ray emission was discovered at the location \citep{anderson2019m81rad}. An issue is that the X-ray observations were performed more than 1 years after the radio bright phase and the X-ray luminosity at these late phases might be low \citep{yu2019aicxray}. Immediate follow-up of radio transients would be helpful to identify their nature.

AIC has been suggested to be a progenitor of FRBs \citep{margalit2019frbaic}. Thus the AIC might be first found as an FRB and an optically-faint radio-bright transient may follow.

Once identified as AIC, our study shows that it is possible to distinguish the evolutionary channel (SD or DD) through the radio LCs, because the two channels predict completely different radio LC evolution as shown in Figs.~\ref{fig:lightcurve_wind} and \ref{fig:lightcurve_const}.

\subsection{43.78+59.3 in M82}
\begin{figure}
\includegraphics[width=\columnwidth]{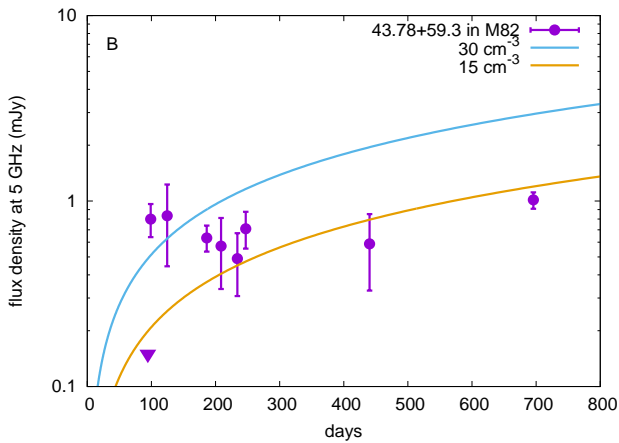}
\caption{
Radio (5~GHz) LC of 43.78+59.3 in M82 from \citet{gendre2013m82} and its comparison with the synthetic radio LC of AIC from the DD channel with the ejecta model B. The triangle shows the observed upper flux limit. The explosion date of the synthetic LCs is 0~days and the time of the observed LC is shifted for comparison.
}
\label{fig:m82}
\end{figure}

\citet{anderson2019m81rad} argue that 43.78+59.3, which is a radio transient appeared in M82 \citep{muxlow2010m82,joseph2011m82,gendre2013m82}, might have similarity to VTC~J095517.5+690813. We here compare our AIC radio LC models with the LC of 43.78+59.3 (Fig.~\ref{fig:m82}). Although the distance to M82 is not well determined, we here assume 3.7~Mpc \citep{marion2015sn2014j}. We take the 5~GHz LC of 43.78+59.3 presented in \citet{gendre2013m82} for the comparison.

Fig.~\ref{fig:m82} shows the comparison between the synthetic AIC radio LC and that of 43.78+59.3. After the quick luminosity increase from the non-detection, 43.78+59.3 had an almost constant luminosity for 600~days. Such a long-term constant luminosity is not expected in AIC from the SD channel (Fig.~\ref{fig:lightcurve_wind}) and we only show models from the DD channel in Fig.~\ref{fig:m82}. Although AIC from the DD channel can have a slower luminosity evolution than that from the SD channel, the LC evolution of 43.78+59.3, especially the initial rapid LC rise followed by the constant luminosity, is not consistent with our AIC radio emission models. Thus, we conclude that 43.78+59.3 is not likely an AIC and has a different origin \citep[e.g.,][]{joseph2011m82}. The long-term LC of 43.78+59.3 enable us to exclude the AIC model. A similar long-term monitoring of VTC~J095517.5+690813 is required to confirm or exclude our AIC model.

\section{Conclusions}
We have investigated the possibility that the recently discovered radio transient in M81, VTC~J095517.5+690813, is a result of AIC. We showed that the radio properties of VTC~J095517.5+690813 are consistent with those expected to be accompained by AIC. If VTC~J095517.5+690813 is an AIC from the SD channel, the CSM density required to account for VTC~J095517.5+690813 is relatively low ($A_\star\simeq 0.01-0.1$). Thus, the donor star should be a main-sequence or He star and the nuclear burning should be stable at the time of AIC. The red giants located at the transient location identified by \textit{HST} are not likely a donor star. If VTC~J095517.5+690813 is an AIC from the DD channel, a relatively high CSM density ($10-100~\mathrm{cm^{-3}}$) is required to explain the radio luminosity. VTC~J095517.5+690813 was not accompanied by an optical transient as predicted for AIC.

Although we find that VTC~J095517.5+690813 can be explained by AIC, it is hard to firmly conclude that it is an AIC. To assess the nature of radio transients, long-term radio transient surveys as well as simultaneous multi-wavelength (including optical and X-ray) transient surveys would be essential.

\section*{Acknowledgements}
The author is supported by the Grants-in-Aid for Scientific Research of the Japan Society for the Promotion of Science (JP17H02864, JP18K13585).




\bibliographystyle{mnras}
\bibliography{mnras} 








\bsp	
\label{lastpage}
\end{document}